\documentclass[pre,twocolumn,10pt,aps,longbibliography]{revtex4-1}

 \usepackage[utf8]{inputenc}
 \usepackage[greek,english]{babel}
 \usepackage{bm}
 \usepackage{verbatim}
 \usepackage{amsmath}
 \usepackage{amssymb}
 \usepackage{amsthm}
 \usepackage{latexsym}
 \usepackage{amsfonts}
 \usepackage{epsfig}
 \usepackage{epstopdf}
 \usepackage{wasysym} 
 \usepackage{subfigure}
 \usepackage{color}
 \definecolor{darkblue}{rgb}{0,0,.5}
 \definecolor{BLUE}{rgb}{0,0,1}
 \definecolor{BLACK}{rgb}{0,0,0}
 \usepackage[linktocpage, colorlinks=true, linkcolor=darkblue, citecolor=darkblue]{hyperref}
 \usepackage[all]{hypcap}
 \usepackage[makeroom]{cancel}
 \usepackage{xfrac}
 \usepackage{lipsum} 
 
\newcommand{\bb}[1]{\textbf{#1}}

\newcommand{\eq}[1]{{Eq.~({#1})}}
\newcommand{\lr}[1]{{\langle {#1}\rangle}}

\begin{document}

\title{Shearing Off the Tree: \\ 
Emerging Branch Structure and Born's Rule in an Equilibrated Multiverse}

\author{Philipp Strasberg}
\author{Joseph Schindler}
\affiliation{F\'isica Te\`orica: Informaci\'o i Fen\`omens Qu\`antics, Departament de F\'isica, Universitat Aut\`onoma de Barcelona, 08193 Bellaterra (Barcelona), Spain}

\date{\today}

\begin{abstract}
Within the many worlds interpretation (MWI) it is believed that, as time passes on, the linearity of the Schr\"odinger equation together with decoherence generate an exponentially growing tree of branches where ``everything happens'', provided the branches are defined for a decohering basis. By studying an example, using exact numerical diagonalization of the Schr\"odinger equation to compute the decoherent histories functional, we find that this picture needs revision. Our example shows decoherence for histories defined at a few times, but a significant fraction (often the vast majority) of branches shows strong interference effects for histories of many times. In a sense made precise below, the histories independently sample an equilibrated quantum process, and, remarkably, we find that only histories that sample frequencies in accordance with Born's rule remain decoherent. Our results suggest that there is more structure in the many worlds tree than previously anticipated, influencing arguments of both proponents and opponents of the MWI.
\end{abstract}

\maketitle

\newtheorem{lemma}{Lemma}[section]
\newtheorem{conj}{Conjecture}


Recent decades have seen intensified research interests in Everett's ``relative state'' formulation of quantum mechanics~\cite{EverettRMP1957}, which became widely known as the many worlds interpretation (MWI)~\cite{DeWittPT1970, DeWittGrahamBook1973, TegmarkSA2003, Carr2007, SaundersEtAlBook2010, WallaceBook2012, Vaidman2021, CarrollPodcast2022}. One reason of its increased popularity is the development of decoherence theory~\cite{ZurekRMP2003, JoosEtAlBook2003}, which can be used to investigate which ``worlds'' behave classical, i.e., are decohered.

\begin{figure}
\centering
\includegraphics[width=.90\linewidth]{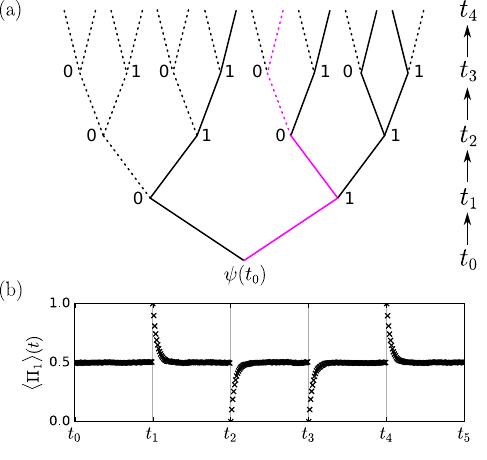}
\caption{(a) Tree structure of the Multiverse for a repeated binary ``0 or 1'' measurement: The standard account posits the reality of all branches (dashed and solid lines), but our results show that only a subset decoheres (solid lines). (b) Trajectory for the probability to obtain outcome `one' conditioned on one branch (pink line in (a)). The numerical parameters of (b) are the same as in Fig.~\ref{fig 5050} with $D=25000$. }
\label{fig MWI}
\end{figure}

Provided one focuses on a basis that decoheres, the many worlds ``Multiverse'' is conventionally pictured as a branching tree with an exponentially growing number of branches as time passes on (we restrict the discussion to nonrelativistic quantum mechanics with a primitive time parameter). This is easily seen for a many worlds description of an idealized frequency experiment, in which a system in a superposition is (re)prepared and measured many times, see Fig.~\ref{fig MWI}(a). Note, however, that a branching could happen for any quantum fluctuation suitably coupled to a macroscopic degree of freedom that decoheres. This is the standard picture of the Multiverse used by both proponents \emph{and} opponents of the MWI~\cite{DeWittPT1970, DeWittGrahamBook1973, TegmarkSA2003, Carr2007, SaundersEtAlBook2010, WallaceBook2012, Vaidman2021, CarrollPodcast2022} and sometimes even \emph{classical} toy-many-worlds-models are used to explain aspects of the MWI.

Yet, the question how long this picture can remain true has not been addressed, it is only clear that decoherence can not proliferate forever in any (effectively) finite dimensional system. Here, we investigate this question in detail by means of a clear example using decoherent histories. We find that decoherence does not stop globally but continues unchanged on a (typically extremely small) subset of branches. This suggests that the many worlds tree has a non-trivial and potentially rich structure.

We continue with the general mathematical formalism and explain how to investigate the decoherence properties of exponentially many branches. Then, we specify a model that can be related to sampling an equilibrated quantum process in an idealized frequency experiment with $L$ independent and repeated trials (however, even if one rejects this interpretation, the numerical results clearly demonstrate the hitherto unobserved phenomenon of a non-trivial (de)coherence structure among the branches). At the end, we discuss the remarkable observation that all decohering branches sample frequencies in accordance with Born's rule.

\bb{Formalism.}
To access the decoherence of branches at many times $L\gg1$ it is natural to use the decoherent histories formalism~\cite{GellMannHartleInBook1990, OmnesRMP1992, HalliwellANY1995, DowkerKentJSP1996, GriffithsBook2001, Griffiths2019}. We are aware of controversies about its use in relation to interpretations of quantum physics, e.g., in Ref.~\cite{VanKampenEtAlPT2000}. In relation to our work, we reject this criticism as we only use the mathematical framework and do \emph{not} postulate decoherent histories (instead, we study their emergence based solely on the Schr\"odinger equation). Seen from that perspective, little controversy remains about the fact that decoherent histories capture an essential aspect of classicality, even though they might not capture all aspects. In particular, various researchers have established a close connection between decoherent histories and the formation of stable records or memories, an important prerequisite to identify classical branches that could support observers~\cite{AlbrechtPRD1992, GellMannHartlePRD1993, FinkelsteinPRD1993, PazZurekPRD1993, HalliwellPRD1999, DoddHalliwellPRD2003, RiedelZurekZwolakPRA2016, HartleArXiv2016}.

Mathematically, we use a complete set of orthogonal projectors $\Pi_x$ associated with some abstract and for simplicity binary property $x\in\{0,1\}$ that decomposes the Hilbert space of an isolated system. The initial state is $|\psi(t_0)\rangle$ and the unitary time evolution operator from $t_k$ to $t_\ell$ is denoted $U_{\ell,k}$. Then, the state compatible with having properties $x_i$ at times $t_i$ for $i\in\{1,2,\dots,L\}$ is 
\begin{equation}\label{eq histories}
 |\psi(\bb x)\rangle = \Pi_{x_L}U_{L,L-1}\cdots\Pi_{x_2}U_{2,1}\Pi_{x_1}U_{1,0}|\psi(t_0)\rangle
\end{equation}
and $\bb x = (x_L,\dots,x_2,x_1)$ is called a \emph{history}. Note that $\bb x$ merely labels the branches $|\psi(\bb x)\rangle$ of a \emph{unitarily} evolving global pure state as per
\begin{equation}\label{eq unitary}
 |\psi(t_L)\rangle = U_{L,0}|\psi(t_0)\rangle = \sum_{\bb x}|\psi(\bb x)\rangle.
\end{equation}
No actual measurement happens in the isolated system. Now, two different histories $\bb x$ and $\bb y$ are decoherent if
\begin{equation}\label{eq decoherent histories}
 \epsilon(\bb x,\bb y) \equiv \frac{|\lr{\psi(\bb y)|\psi(\bb x)}|}{\sqrt{\lr{\psi(\bb x)|\psi(\bb x)}}\sqrt{\lr{\psi(\bb y)|\psi(\bb y)}}} \ll 1.
\end{equation}
Note that $\epsilon(\bb x,\bb y)\le 1$ by Cauchy-Schwarz' inequality.

Unfortunately, the $2^L$ many relative states $|\psi(\bb x)\rangle$ make the evaluation of \eq{\ref{eq decoherent histories}} for $L\gg1$ practically impossible. Therefore, we focus our attention on a specific question (whose motivation will become clear below) and introduce the relative state 
\begin{equation}
 |\psi(m)\rangle \equiv \sum_{\bb x} \delta_{m,n_1(\bb x)} |\psi(\bb x)\rangle, ~~~ m\in\{0,1,\dots,L\}.
\end{equation}
Here, $\delta_{m,n}$ is the Kronecker symbol and $n_1(\bb x) \equiv \sum_{i=1}^L x_i$ the \emph{net} number of `ones' in $\bb x$. These states still satisfy $\sum_m|\psi(m)\rangle = |\psi(t_L)\rangle$ and $\sum_m \lr{\psi(m)|\psi(m)} = 1$. They therefore label a complete set of branches in which potential observers care only about the relative frequency $m/L$ of `ones' but forgot the information about which particular sequence $\bb x$ was realized.

\bb{Toy model.}
Within this binary setting \emph{any} Hamiltonian can be written as
\begin{equation}
 H = \left(\begin{array}{c|c}
            H_{00} & H_{01} \\ \hline
            H_{10} & H_{11} \\
           \end{array}\right)
\end{equation}
with $H_{ij} = \Pi_i H\Pi_j$. For reasons we explain below, we take $H_{00}$ ($H_{11}$) to be diagonal matrices with $d_0$ ($d_1$) evenly spaced eigenenergies in the interval $[0,\delta\epsilon]$ (in the simulation we set $\delta\epsilon=0.5$) such that the total Hilbert space dimension is $D = d_0 + d_1$. These subspaces interact via $H_{01} = H_{10}^\dagger = \lambda R$, where $R$ is a random matrix uniformly filled with zero-mean-unit-variance Gaussian random numbers and $\lambda$ is a coupling strength. We will restrict the discussion now to the relevant weak coupling regime, described by the condition $c \equiv 8\lambda^2d_0d_1/(D\delta\epsilon^2) \ll1$, but we later also study numerically the strong coupling regime.

We first note about this model that decoherence for small $L$ is established~\cite{GemmerSteinigewegPRE2014, StrasbergSP2023, StrasbergReinhardSchindlerArXiv2023}. In addition, within an open system approach, $\Pi_0$ and $\Pi_1$ can be seen as projectors onto pointer states of a two-level system weakly coupled to an environment of dimension $D/2$. Environmentally induced decoherence has been also established in this case (e.g., in Refs.~\cite{GorinEtAlNJP2008, GenwayHoLeePRL2013, AlbrechtBaunachArrasmithPRD2022, YanZurekNJP2022}), but it is not necessary to refer to any system-environment tensor product structure in the following. Moreover, equilibration and thermalization of this model is also well established (e.g., in Refs.~\cite{PereyraJSP1991, EspositoGaspardPRE2003b, LebowitzPasturJPA2004, BartschSteinigewegGemmerPRE2008, RieraCampenySanperaStrasbergPRXQ2021}) and the characteristic relaxation time scale is well described by $\tau = \delta\epsilon/(2\pi\lambda^2D)$~\cite{BartschSteinigewegGemmerPRE2008}.

This toy model can therefore be regarded as an archetype model behaving in agreement with decoherence theory and statistical mechanics. Moreover, the success of random matrix theory to describe \emph{generic} properties of complex systems~\cite{Wigner1967, BrodyEtAlRMP1981, BeenakkerRMP1997, GuhrMuellerGroelingWeidenmuellerPR1998, HaakeBook2010, DAlessioEtAlAP2016, DeutschRPP2018} motivates the conjecture that it also captures relevant aspects of realistic systems, in particular owing to the close connection between random matrix theory and realistic quantum many-body systems (eigenstate thermalization hypothesis~\cite{DeutschPRA1991, SrednickiPRE1994, SrednickiJPA1999, DAlessioEtAlAP2016, DeutschRPP2018, ReimannDabelowPRE2021}).

Finally, we choose a Haar random initial state $|\psi(t_0)\rangle$ and times $t_{j+1}-t_j \gg \tau$ (numerically, we randomly sample $t_{j+1}-t_j\in[19.5\tau,20.5\tau]$). The latter choice ensures that the projectors $\Pi_x$ reach their equilibrium value $\lr{\Pi_x} \approx d_x/D$ (with exponentially suppressed fluctuations around its mean~\cite{FarquharLandsbergPRSLA1957, BocchieriLoingerPR1958, GemmerMichelMahlerBook2004, PopescuShortWinterNatPhys2006, HeitmanEtAlZFN2020}) before \emph{each} $t_j$, as shown in Fig.~\ref{fig MWI}(b). Thus, from a macroscopic point of view the systems looks identically prepared at each $t_j$, although subtle, hidden microscopic correlations remain, of course.

These considerations motivate the picture of an \emph{idealized frequency experiments} of an equilibrated quantum process, where $|\psi(m)\rangle$ is associated with a record of $m$ `ones' after $L$ independent trials---provided the branch decoheres to allow the formation of a record. These statements will be precisely quantified below. Here, we only remark that, of course, our toy model is unable to capture any realistic experimental setup owing to obvious numerical limitations. We use the convenient metaphor of a frequency experiments because we capture the key aspect of having a binary degree of freedom (which, as noted above, could label the states of an open quantum system coupled to a complicated laboratory environment) probed and (re)prepared many times in a seemingly independent way. Moreover, our main point---the emergence of a non-trivial and non-classical branch structure of the many-worlds tree---does not rely on this purported connection.

In particular, we like to counteract sceptical voices announcing that the random Hamiltonian or the random initial state introduce classical noise or probabilities from the outside into our treatment. This is wrong because all presented numerical results are obtained for a \emph{single} choice of $H$ (kept fixed throughout the dynamics) and a \emph{single} choice of $|\psi(t_0)\rangle$, we do not perform any averages. The global state remains pure at all times and all uncertainties are entirely of quantum origin. Nevertheless, and very importantly, we have observed the results to be generic, i.e., different choices for $H$ and $|\psi(t_0)\rangle$ give rise to similar behaviour. Backed up by the success of random matrix theory~\cite{Wigner1967, BrodyEtAlRMP1981, BeenakkerRMP1997, GuhrMuellerGroelingWeidenmuellerPR1998, HaakeBook2010, DAlessioEtAlAP2016, DeutschRPP2018} and typicality~\cite{GemmerMichelMahlerBook2004, HeitmanEtAlZFN2020}, this motivates the conclusion that our results are more widely applicable.

\begin{figure}[t]
 \centering\includegraphics[width=.98\linewidth]{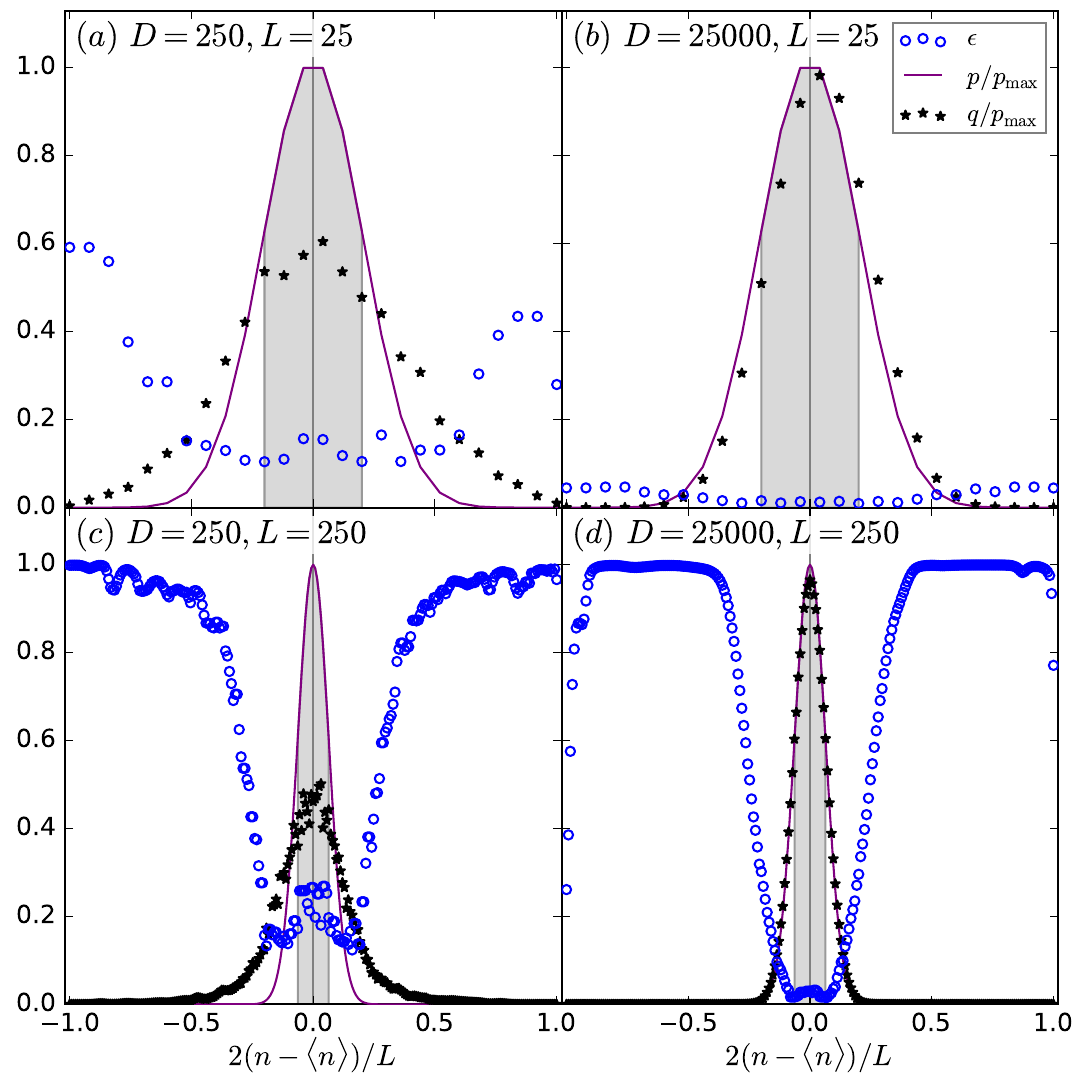}
 \caption{Coherence measure $\epsilon(n)$ (blue circles) and probabilities $p(n)$ (solid purple line) and $q(n)$ (black stars) as a function of $n$. Here and in all figures: $\lr{n} = Ld_1/D$ is the expected number of ones according to $p(n)$, we always rescale $p(n)$ and $q(n)$ by dividing by $p_\text{max} = \max_n p(n)$, the $x$ axis is rescaled to display $n$ on an interval of size two, and the gray area corresponds to one standard deviation of $p(n)$. }
 \label{fig 5050}
\end{figure}

\bb{Numerical evidence.}
In all plots we quantify the amount of (de)coherence by considering
\begin{equation}
 \epsilon(n) = \max_{m\neq n}\frac{|\lr{\psi(m)|\psi(n)}|}{\sqrt{\lr{\psi(m)|\psi(m)}}\sqrt{\lr{\psi(n)|\psi(n)}}} \in[0,1].
\end{equation}
If $\epsilon(n)$ is close to zero, the branch giving rise to $n$ `ones' is decohered from all other branches and allows the formation of stable, classical records; whereas $\epsilon(n)$ close to one implies strong interference with other branches $m\neq n$, preventing a formation of reliable records. Note that strong coherence between the branches $|\psi(n)\rangle$ implies also strong coherence between the fine-grained branches $|\psi(\bb x)\rangle$ as shown in the Appendix.

Moreover, we also compare the two probabilities
\begin{equation}\label{eq probabilities}
 q(n) \equiv \lr{\psi(n)|\psi(n)}, ~~~ p(n) \equiv \binom{L}{n} p_1^n (1-p_1)^{L-n},
\end{equation}
where $q(n)$ is the exact probability to observe $n$ times `one', whereas $p(n)$ is obtained by applying Born's rule with a single-time probability $p_1$ to observe $x=1$ for $L$ independent trials. Thus, $p(n)$ is a binomial distribution and from what we said above we expect (and verify below) that $p_1 = d_1/D$.

We start with equal subspace dimensions $d_0 = d_1$ and ensure weak coupling by setting $c=0.0025$. The time evolution of a single history along a particular branch for a few steps then looks as in Fig.~\ref{fig MWI}~(b) in unison with our expectations. Turning to longer histories, Fig.~\ref{fig 5050} displays $\epsilon(n)$, $q(n)$ and $p(n)$. First, in Figs.~\ref{fig 5050}(a) and~(b) we consider a history of length $L=25$ and compare two system sizes: $D=250$ and $D=25000$. We confirm that decoherence is much stronger for larger system size, in unison with the scaling laws of  Refs.~\cite{StrasbergEtAlPRA2023, StrasbergSP2023, StrasbergReinhardSchindlerArXiv2023}. Moreover, we have $q(n)\approx p(n)$ for $D=25000$, but see significant deviations for $D=250$ owing to the strong influence of finite size effects. Thus, we see probabilities $q(n)$ emerging that start to behave as in an ideal frequency experiment for large $D$. However, the coherences $\epsilon(n)$ do not yet show any clear pattern, but $L=25$ is still quite short.

Things start to change drastically for long histories with $L=250$ as shown in Figs.~\ref{fig 5050}(c) and~(d). Suddenly, there is a very clear minimum around $n\approx\lr{n}$. Remarkably, away from it $\epsilon(n)$ shoots up close to its \emph{maximum value} for both $D=250$ and $D=25000$, signifying the strongest possible coherences. Moreover, we see now that $q(n)$ perfectly fits $p(n)$ for $D=25000$, whereas strong deviations continue to exist for $D=250$ (as expected). Thus, for $D=25000$ we are close to an ideal frequency experiment, and coherent effects start growing approximately when $|n-\lr{n}|$ exceeds one standard deviation.

\begin{figure}[t]
 \centering\includegraphics[width=.98\linewidth]{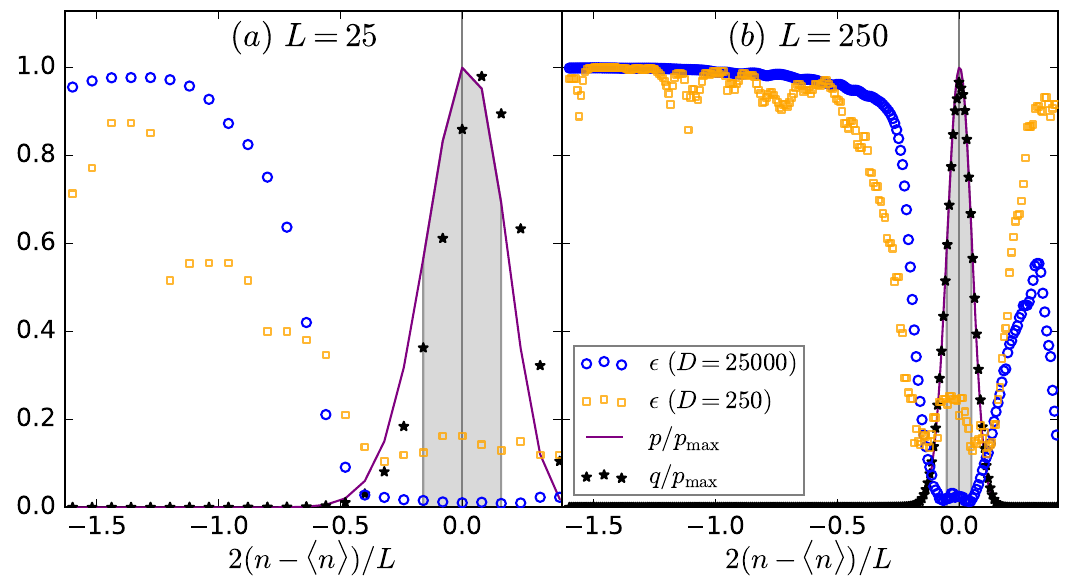}
 \caption{Coherence measure $\epsilon(n)$ for $D=250$ (orange squares) and $D=25000$ (blue circles) and probabilities $p(n)$ and $q(n)$ (as in Fig.~\ref{fig 5050}) for $D=25000$ as a function of $n$. }
 \label{fig 2080}
\end{figure}

However, the 50/50 splitting of the subspaces is special, and to challenge our approach we continue by considering a 20/80 splitting with $d_1=4d_0$, implying $p_1 = 0.8$. The results are shown in Fig.~\ref{fig 2080}, where we display $\epsilon(n)$ for $D=250$ and $D=25000$ jointly in one plot, whereas we plot $q(n)$ and $p(n)$ only for $D=25000$. In Fig.~\ref{fig 2080}(a) we can now see some clear signatures for large coherences $\epsilon(n)$ if $n$ deviates significantly from $\lr{n}$ even for short histories of length $L=25$. Moreover, in both Figs.~\ref{fig 2080}(a) and~(b) we see that $\epsilon(n)$ for $D=25000$ can be larger than $\epsilon(n)$ for $D=250$, contrary to what one would expect from the scaling behaviour for small $L$~\cite{StrasbergEtAlPRA2023, StrasbergSP2023, StrasbergReinhardSchindlerArXiv2023}. Apart from these additional observations, the evidence in Fig.~\ref{fig 2080} matches the conclusions from Fig.~\ref{fig 5050}.

\begin{figure}[t]
 \centering\includegraphics[width=.98\linewidth]{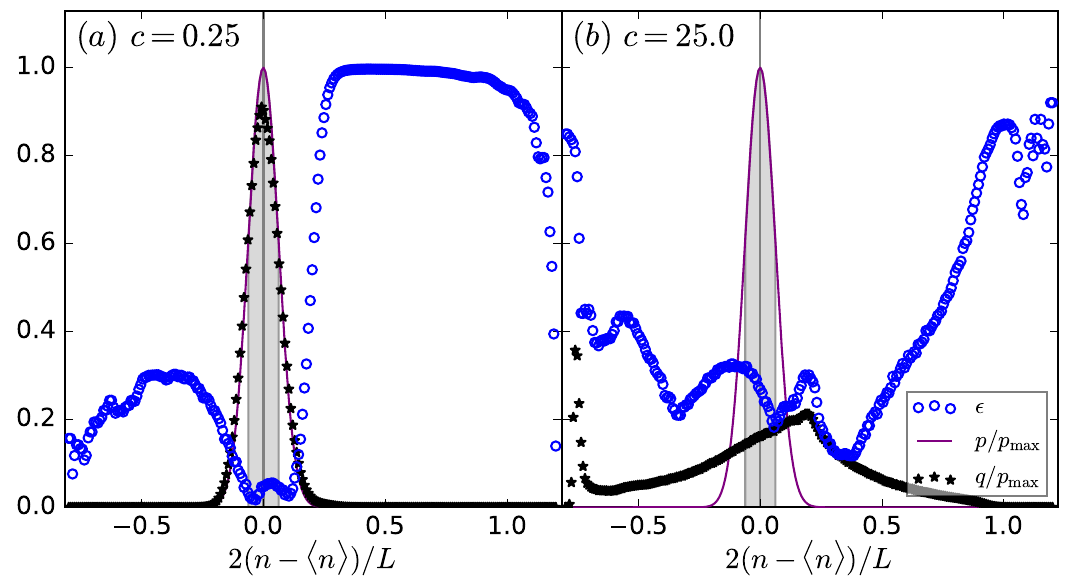}
 \caption{Plot of $\epsilon(n)$, $p(n)$ and $q(n)$ for $L=250$ and $D=25000$ (legend as in Fig.~\ref{fig 5050}). For (b) we rescaled $\tau$ to $10\tau$ to ensure equilibration between different trials.}
 \label{fig 6040fast}
\end{figure}

Finally, we consider the strong coupling case as a counterexample. Indeed, decoherent histories are expected to emerge only for slow and coarse observables of a large dimensional system~\cite{VanKampenPhys1954, GellMannHartlePRD1993, HalliwellPRL1999, StrasbergSP2023, StrasbergReinhardSchindlerArXiv2023, StrasbergEtAlPRA2023}, but at strong coupling the observable is no longer slow. The numerical results for $L=250$ and $D=25000$ with $d_0=3d_1/2$ are shown in Fig.~\ref{fig 6040fast}~(a) for $c=0.25$ (moderate coupling) and (b) for $c=25.0$ (very strong coupling). Figure~\ref{fig 6040fast}(a) still admits an interpretation along the lines of the previous figures, showing a certain robustness of our results. Nevertheless, deviations between $q(n)$ and $p(n)$ become noticably visible, in particular around $n\approx\lr{n}$. The picture drastically changes for very strong coupling in Fig.~\ref{fig 6040fast}(b). While $\epsilon(n)$ still shows some non-trivial structure, it seems hard to relate it to any physical properties. Moreover, $q(n)$ and $p(n)$ are completely distinct, indicating strong correlations between the trials.

Before turning to a broader discussion, it is worth to summarize our numerical results within the MWI. First, for $L\gg1$ the decohering branches of the universal wave function of our toy model are only those sequences where $m/L\approx p_1$. Observers \emph{inside} it (which, admittably, are hard to picture here) and without prior knowledge of Born's rule would therefore (self)locate themselves in worlds obeying Born's rule, i.e., they could \emph{infer} the validity of the purple solid line in Fig.~\ref{fig 5050} and~\ref{fig 2080}. Moreover, the fact that $p(n)\approx q(n)$ (which is something an observer inside the Multiverse could not confirm) proves that the sampled frequencies correspond to \emph{independent} trials.

\bb{Discussion.}
To the best of our knowledge, this is the first study investigating decoherence properties for $L\gg1$. In contrast to the widely held believe that single-time decoherence proliferates together with the branches, we have shown that the many worlds tree possesses a non-trivial structure charactrized by branches that show the \emph{highest possible} degree of coherence, i.e., $\epsilon(n)\approx1$. 

Moreover, we established a connection to another important result. If one consider a unitary protocol where $L$ \emph{decorrelated} subsystems interact with a measurement apparatus (either in parallel at the same time or sequentially), then it can be shown that for any fixed $\delta>0$
\begin{equation}\label{eq Born concentration}
 \lim_{L\rightarrow\infty} \left\|\Pi_\text{Born}^{(L)}(\delta)|\psi(t_f)\rangle - |\psi(t_f)\rangle\right\| = 0,
\end{equation}
where $\Pi_\text{Born}^{(L)}(\delta)$ projects the final universal wave function $|\psi(t_f)\rangle$ onto a subspace compatible with Born's rule up to an error $\delta$. This result is in essence the law of large numbers and was already used by Everett~\cite{EverettRMP1957} (see also Refs.~\cite{SaundersInBook2010, AguirreTegmarkPRD2011, LazaroviciQR2023}), but it relies crucially on the assumed independence of the measured systems. Obviously, if the Universe is only a single wave function, it is full of microscopic correlations that have formed during its evolution or were already present initially. Here, we showed that Eq.~(\ref{eq Born concentration}) emerges also in our equilibrated model \emph{without} assuming independence or decorrelated subsystems. This follows from $q(n)\approx p(n)$ and the concentration of $p(n)$ around its mean, which clearly calls for a rigorous analytical result in the future. 

Whereas the previous two paragraphs summarized two unambiguous main results of our study, a remaining big question is whether the structure of (de)coherence together with Eq.~(\ref{eq Born concentration}) can be used to explain Born's rule within the MWI \emph{in general}. The short and honest answer is: We are sceptical, but we do not know!

To put this question in context, we recall the attempts to derive Born's rule within the MWI~\cite{EverettRMP1957, DeWittGrahamBook1973, VaidmanISPS1998, DeutschPRSCA1999, WallaceSHPSB2003, GreavesSHPSB2004, ZurekPRA2005, SaundersInBook2010, SaundersEtAlBook2010, AguirreTegmarkPRD2011, WallaceBook2012, SebensCarrollBJP2018, MasanesGalleyMuellerNC2019, McQueenVaidmanSHPSB2019, Vaidman2020, SaundersPRSA2021, Vaidman2021, ZurekEnt2022, ShortQuantum2023, LazaroviciQR2023}, but all rely on additional postulates and none could convince its opponents~\cite{SaundersEtAlBook2010, MaudlinBook2019}---not to mention that even the proposed \emph{solutions} show a remarkable lack of consensus among themselves. In particular, the so-called theory confirmation problem asks why do we find ourselves in a world compatible with Born's rule if the \emph{vast majority} of branches is incompatible with it? Indeed, the number of worlds is, by simple counting, always given by a binomial coefficient $\binom{L}{n}$ centered around $\lr{n} = L/2$. Thus, for $p_1\in(0,1)\setminus\{1/2\}$ the term in Eq.~(\ref{eq Born concentration}), while small with respect to the Hilbert space norm, contains the majority of worlds and, as Kent emphasizes~\cite{KentInBook2010}, \emph{``It's no more scientifically respectable to declare that we can [...] confirm Everettian quantum theory by neglecting the observations made on selected low Born weight branches [...] unless we add further structure to the theory [...].''}

In our toy model, we clearly see such further structure emerging from the Schr\"odinger equation itself, i.e., we find an unambiguous resolution of the quantum measurement problem for an arguably irrelevant model of the Multiverse. Unfortunately, there is no evidence that this could solve the theory confirmation problem in general, but in light of the variety of attempts we believe this novel direction deserves attention.

Finally, we try to give some physical intuition for our results. Recall that decoherence requires coarse-graining to ``hide'' coherences in inaccessible microscopic degrees of freedom, but if these microscopic degrees of freedom are restricted to evolve in a small and non-generic subspace, they no longer can induce decoherence effectively. We believe that we observe this effect here: as $L$ becomes large, some sequences restrict the relative state on this branch to a very small and non-generic subspace, thus preventing decoherence. This effect might be related to a recent result in pure state statistical mechanics~\cite{DowlingEtAlQuantum2023, DowlingEtAlSPC2023}, where the authors showed that the distinguishability between an arbitrary quantum process sampled at $L$ random times from a corresponding equilibrium process is bounded by a number that scales as $2^{2L}/D$ in the worst case, thus revealing a competition between $L$ and $D$. However, the effect seems subtle as already Fig.~\ref{fig 2080} shows that increasing the Hilbert space dimension does not necessarily decrease the coherence among all branches. Moreover, it does not matter whether we humans are able to practically use the information contained in a sequence $\bb x$ to infer the true relative state. Within the realist stance behind the MWI it is a matter of principle whether the relative state compatible with \emph{all} the information out there looks generic or not.

\bb{Concluding perspectives.}
We demonstrated that the many worlds tree can have a non-trivial structure conflicting with the naive branch realism that is found behind many arguments of proponents \emph{and} opponents of the MWI alike. We further showed that Eq.~(\ref{eq Born concentration}) holds in a unitarily evolving and quantum correlated Universe. Finally, we observed the \emph{emergence} of Born's rule (which, interestingly, is also a feature of de Broglie-Bohm theory~\cite{ValentiniWestmanPRSA2005, ValentiniInBook2010}) within our toy model, but we have no evidence that this is true in general. 

While many open questions remain, the present approach demonstrates that \emph{fundamental} aspects of the MWI can be studied by using nothing but Schr\"odinger's equation (in a non-relativstic context) \emph{without} approximations or additional metaphysical postulates. We have the tools to rigorously access the decoherence of (long) histories and the structure of the (potential) quantum mechanical Multiverse (see also Ref.~\cite{StrasbergReinhardSchindlerArXiv2023}), and it might be full of marvelous wonders. Whether they speak in favour or against the MWI needs to be found out.

\bb{Acknowledgements.}
We gratefully acknowledge discussions with Teresa E.~Reinhard and Giulio Gasbarri. Finanical support by MICINN with funding from European Union NextGenerationEU (PRTR-C17.I1) and by the Generalitat de Catalunya (project 2017-SGR-1127) are acknowledged. PS is further supported by ``la Caixa'' Foundation (ID 100010434, fellowship code LCF/BQ/PR21/11840014), the European Commission QuantERA grant ExTRaQT (Spanish MICIN project PCI2022-132965), and the Spanish MINECO (project PID2019-107609GB-I00) with the support of FEDER funds.


\bibliography{/home/philipp/Documents/references/books,/home/philipp/Documents/references/open_systems,/home/philipp/Documents/references/thermo,/home/philipp/Documents/references/info_thermo,/home/philipp/Documents/references/general_QM,/home/philipp/Documents/references/math_phys,/home/philipp/Documents/references/equilibration,/home/philipp/Documents/references/time,/home/philipp/Documents/references/cosmology,/home/philipp/Documents/references/general_refs}

\onecolumngrid
\appendix
\section*{}

\begin{center}
 \textbf{Appendix}
\end{center}

We show that
\begin{equation}\label{eq lemma}
 \epsilon(m,n)^2 = \frac{|\lr{\psi(m)|\psi(n)}|^2}{\lr{\psi(m)|\psi(m)}\lr{\psi(n)|\psi(n)}} = 1
 ~~~ \Rightarrow ~~~
 \epsilon(\bb x,\bb y)^2
 = \frac{|\lr{\psi(\bb x)|\psi(\bb y)}|^2}{\lr{\psi(\bb x)|\psi(\bb x)}\lr{\psi(\bb y)|\psi(\bb y)}} = 1
\end{equation}
for all $\bb x$ and $\bb y$ such that $n_1(\bb x)\neq n_1(\bb y)$. If instead $m \equiv n_1(\bb x) = n_1(\bb y)$, it follows by definition that $\epsilon(m,m)=1$. Moreover, we also assume that $\lr{\psi(\bb x)|\psi(\bb x)} > 0$ and $\lr{\psi(\bb y)|\psi(\bb y)} > 0$, i.e., we exclude all branches that do not exist. Note that in our example all branches have non-zero weight (even though the weight might be extremely small).

To verify \eq{\ref{eq lemma}}, we note that $\epsilon(m,n)^2 = 1$ is equivalent to
\begin{equation}
 0 = \sum_{\bb x,\bb x',\bb y,\bb y'} A(\bb x,\bb x',\bb y,\bb y',m,n) [\epsilon(\bb x,\bb y)\epsilon(\bb y',\bb x') - \epsilon(\bb x,\bb x') \epsilon(\bb y',\bb y)],
\end{equation}
where
\begin{equation}
 A(\bb x,\bb x',\bb y,\bb y',m,n) \equiv \lr{\psi(\bb x)|\psi(\bb x)} \lr{\psi(\bb x')|\psi(\bb x')} \lr{\psi(\bb y)|\psi(\bb y)} \lr{\psi(\bb y')|\psi(\bb y')} \delta_{m,n_1(\bb x)} \delta_{n,n_1(\bb y)} \delta_{m,n_1(\bb x')} \delta_{n,n_1(\bb y')}
\end{equation}
Since $A(\bb x,\bb x',\bb y,\bb y',m,n) > 0$ if $m \equiv n_1(\bb x) = n_1(\bb x') \neq n \equiv n_1(\bb y) = n_1(\bb y')$, we find that the condition $\epsilon(m,n)^2=1$ for $m\neq n$ implies
\begin{equation}
 \epsilon(\bb x,\bb y)\epsilon(\bb y',\bb x') = \epsilon(\bb x,\bb x') \epsilon(\bb y',\bb y).
\end{equation}
We now choose $\bb x = \bb x'$ and $\bb y=\bb y'$ with $n_1(\bb x) = m \neq n_1(\bb y) = n$, which implies $\epsilon(\bb x,\bb y) = 1$ since $\epsilon(\bb x,\bb x) = 1$ by definition.

\end{document}